\documentstyle[12pt]{article}
\begin{document}

\def\ds{\displaystyle}
\def\beq{\begin{equation}}
\def\eeq{\end{equation}}
\def\bea{\begin{eqnarray}}
\def\eea{\end{eqnarray}}
\def\beeq{\begin{eqnarray}}
\def\eeeq{\end{eqnarray}}
\def\ve{\vert}
\def\vel{\left|}
\def\ver{\right|}
\def\nnb{\nonumber}
\def\ga{\left(}
\def\dr{\right)}
\def\aga{\left\{}
\def\adr{\right\}}
\def\lla{\left<}
\def\rra{\right>}
\def\rar{\rightarrow}
\def\nnb{\nonumber}
\def\la{\langle}
\def\ra{\rangle}
\def\ba{\begin{array}}
\def\ea{\end{array}}
\def\tr{\mbox{Tr}}
\def\ssp{{\Sigma^{*+}}}
\def\sso{{\Sigma^{*0}}}
\def\ssm{{\Sigma^{*-}}}
\def\xis0{{\Xi^{*0}}}
\def\xism{{\Xi^{*-}}}
\def\qs{\la \bar s s \ra}
\def\qu{\la \bar u u \ra}
\def\qd{\la \bar d d \ra}
\def\qq{\la \bar q q \ra}
\def\gGgG{\la g^2 G^2 \ra}
\def\q{\gamma_5 \not\!q}
\def\x{\gamma_5 \not\!x}
\def\g5{\gamma_5}
\def\sb{S_Q^{cf}}
\def\sd{S_d^{be}}
\def\su{S_u^{ad}}
\def\ss{S_s^{??}}
\def\sbp{{S}_Q^{'cf}}
\def\sdp{{S}_d^{'be}}
\def\sup{{S}_u^{'ad}}
\def\ssp{{S}_s^{'??}}
\def\sig{\sigma_{\mu \nu} \gamma_5 p^\mu q^\nu}
\def\fo{f_0(\frac{s_0}{M^2})}
\def\ffi{f_1(\frac{s_0}{M^2})}
\def\fii{f_2(\frac{s_0}{M^2})}
\def\O{{\cal O}}
\def\sl{{\Sigma^0 \Lambda}}
\def\es{\!\!\! &=& \!\!\!}
\def\ar{&+& \!\!\!}
\def\ek{&-& \!\!\!}
\def\cp{&\times& \!\!\!}
\def\se{\!\!\! &\simeq& \!\!\!}
\def\kpm{&\pm& \!\!\!}
\def\kmp{&\mp& \!\!\!}


\def\simlt{\stackrel{<}{{}_\sim}}
\def\simgt{\stackrel{>}{{}_\sim}}


\renewcommand{\textfraction}{0.2}    
\renewcommand{\topfraction}{0.8}   

\renewcommand{\bottomfraction}{0.4}   
\renewcommand{\floatpagefraction}{0.8}
\newcommand\mysection{\setcounter{equation}{0}\section}

\def\baeq{\begin{appeq}}     \def\eaeq{\end{appeq}}  
\def\baeeq{\begin{appeeq}}   \def\eaeeq{\end{appeeq}}
\newenvironment{appeq}{\beq}{\eeq}   
\newenvironment{appeeq}{\beeq}{\eeeq}
\def\bAPP#1#2{
 \markright{APPENDIX #1}
 \addcontentsline{toc}{section}{Appendix #1: #2}
 \medskip
 \medskip
 \begin{center}      {\bf\LARGE Appendix #1 :}{\quad\Large\bf #2}
\end{center}
 \renewcommand{\thesection}{#1.\arabic{section}}
\setcounter{equation}{0}
        \renewcommand{\thehran}{#1.\arabic{hran}}
\renewenvironment{appeq}
  {  \renewcommand{\theequation}{#1.\arabic{equation}}
     \beq
  }{\eeq}
\renewenvironment{appeeq}
  {  \renewcommand{\theequation}{#1.\arabic{equation}}
     \beeq
  }{\eeeq}
\nopagebreak \noindent}

\def\eAPP{\renewcommand{\thehran}{\thesection.\arabic{hran}}}

\renewcommand{\theequation}{\arabic{equation}}
\newcounter{hran}
\renewcommand{\thehran}{\thesection.\arabic{hran}}

\def\bmini{\setcounter{hran}{\value{equation}}
\refstepcounter{hran}\setcounter{equation}{0}
\renewcommand{\theequation}{\thehran\alph{equation}}\begin{eqnarray}}
\def\bminiG#1{\setcounter{hran}{\value{equation}}
\refstepcounter{hran}\setcounter{equation}{-1}
\renewcommand{\theequation}{\thehran\alph{equation}}
\refstepcounter{equation}\label{#1}\begin{eqnarray}}


\newskip\humongous \humongous=0pt plus 1000pt minus 1000pt
\def\caja{\mathsurround=0pt}


\title{
         {\Large
                 {\bf
New physics effects in $\Lambda_b \rar \Lambda \ell^+ \ell^-$ decay 
with lepton polarizations
                 }
         }
      }

\author{\vspace{1cm}\\
{\small T. M. Aliev$^a$ \thanks
{e-mail: taliev@metu.edu.tr}\,\,,
A. \"{O}zpineci$^b$ \thanks
{e-mail: ozpineci@ictp.trieste.it}\,\,,
M. Savc{\i}$^a$ \thanks
{e-mail: savci@metu.edu.tr}} \\
{\small a Physics Department, Middle East Technical University, 
06531 Ankara, Turkey}\\
{\small b  The Abdus Salam International Center for Theoretical Physics,
I-34100, Trieste, Italy} }
\date{}

\begin{titlepage}
\maketitle
\thispagestyle{empty}

\begin{abstract}
We study lepton polarization asymmetries in the $\Lambda_b \rar \Lambda
\ell^+ \ell^-$ decay using the most general model independent effective
Hamiltonian. The dependence of the lepton polarizations and their
combinations on the new Wilson coefficients are studied in detail.
It is observed that there is a region for the new Wilson coefficients 
for which the branching ratio coincides with the standard model 
prediction, while lepton polarizations show considerable departure 
from standard model. 
\end{abstract}

~~~PACS numbers: 12.60.--i, 13.30.--a, 14.20.Mr
\end{titlepage}

\section{Introduction}
Flavor--changing neutral current (FCNC) $b \rar s(d) \ell^+ \ell^-$ decays
provide important tests for the gauge structure of the standard model (SM)
at one--loop level. Moreover, $b \rar s(d) \ell^+ \ell^-$ decay is also very
sensitive to the new physics beyond SM.  
New physics effects manifest themselves in rare decays
in two different ways, either through new combinations to the new Wilson
coefficients or through the new operator structure
in the effective Hamiltonian which is absent in the SM. One
of the efficient ways in establishing new physics beyond the SM is the
measurement of the lepton polarization \cite{R4621}--\cite{R4628}.

In this paper we investigate the possibility of searching for new physics
in the heavy baryon decays $\Lambda_b \rar \Lambda
\ell^+ \ell^-$ using the most general model independent form
of effective Hamiltonian. 

The main problem for the description of exclusive decays is to
evaluate the form factors, i.e., matrix elements
of the effective Hamiltonian between initial and final hadron states. It
is well known that for describing baryonic $\Lambda_b \rar \Lambda
\ell^+ \ell^-$ decay quite some number of form factors are needed
(see for example \cite{R4629}). However, when heavy quark effective theory
(HQET) is applied only two independent form factors appear \cite{R46210}. 

It should be mentioned here that the exclusive decay
$\Lambda_b \rar  \Lambda \ell^+ \ell^-$ decay is studied
in the SM, the two Higgs doublet model and using the general form
of the effective Hamiltonian, in \cite{R4629}, \cite{R46211} and 
\cite{R46212}, respectively.    

The sensitivity of the lepton polarizations to the new Wilson coefficients
in the $B \rar K^\ast \ell^+ \ell^-$ decay is investigated
in \cite{R46213} using the general form of the Hamiltonian. 
It is shown in this work that the lepton
polarizations are very sensitive to the scalar and tensor interactions. In
this connection it is natural to ask to which new Wilson coefficients 
the lepton polarizations are strongly sensitive, in the "heavy baryon
$\rar$ light baryon $\ell^+ \ell^-$" decays. In the present work we try to
answer this question.

The paper is organized as follows. In section
2, using the most general form of the four--Fermi interaction, the general
expressions for the longitudinal, transversal and normal polarizations
of leptons are derived. In section 3 we investigate the sensitivity
of these polarizations, as well as combined polarizations of
lepton and antilepton to the new Wilson coefficients.

\section{Lepton polarizations}

In order to calculate lepton polarization in $\Lambda_b \rar \Lambda\ell^+
\ell^-$ decay, we start with the effective Hamiltonian for the $b \rar
\ell^+ \ell^-$ transition. This effective Hamiltonian can be written
in terms of twelve model independent four--Fermi interactions \cite{R4628},

\bea
\label{e1}
\lefteqn{
{\cal M} = \frac{G \alpha}{4\sqrt{2} \pi} V_{tb}V_{ts}^\ast \Bigg\{
C_{SL} \bar s i \sigma_{\mu\nu} \frac{q^\nu}{q^2} L b \bar\ell
\gamma_\mu \ell +
C_{BR} \bar s i \sigma_{\mu\nu} \frac{q^\nu}{q^2} b \bar\ell \gamma_\mu
\ell +
C_{LL}^{tot} \bar s_L \gamma^\mu b_L \bar \ell_L\gamma_\mu \ell_L} \nnb\\
\ar C_{LR}^{tot} \bar s_L \gamma^\mu b_L \bar \ell_R \gamma_\mu \ell_R +
C_{RL} \bar s_R \gamma^\mu b_R \bar \ell_L \gamma_\mu \ell_L +
C_{RR} \bar s_R \gamma^\mu b_R \bar \ell_R \gamma_\mu \ell_R \nnb \\
\ar C_{LRLR} \bar s_L b_R \bar \ell_L \ell_R +
C_{RLLR} \bar s_R b_L \bar \ell_L \ell_R +
C_{LRRL} \bar s_L b_R \bar \ell_R \ell_L +
C_{RLRL} \bar s_R b_L \bar \ell_R \ell_L \nnb \\
\ar C_T \bar s \sigma^{\mu\nu} b \bar \ell \sigma_{\mu\nu} \ell +
i C_{TE} \epsilon^{\mu\nu\alpha\beta} \bar s \sigma_{\mu\nu} s
\sigma_{\alpha\beta} \ell \Bigg\}~,
\eea
where $L=(1-\gamma_5)/2$ and $R=(1+\gamma_5)/2$ are the chiral operators.
The coefficients of the first two terms, $C_{SL}$ and $C_{BR}$ are the
nonlocal Fermi interactions, which correspond to $-2 m_s C_7^{eff}$ and $-2
m_b C_7^{eff}$ in the SM, respectively. The next four terms with coefficients 
$C_{LL}^{tot},~C_{LR}^{tot},~ C_{RL}$ and $C_{RR}$ in Eq. (\ref{e1}) 
describe vector type interactions. Two of these coefficients
$C_{LL}^{tot}$ and $C_{LR}^{tot}$ contain SM results in the form
$C_9^{eff}-C_{10}$ and $C_9^{eff}-C_{10}$, respectively. For this reason
we can write \bea
\label{e2}
C_{LL}^{tot} \es C_9^{eff}- C_{10} + C_{LL}~, \nnb \\
C_{LR}^{tot} \es C_9^{eff}+ C_{10} + C_{LL}~,
\eea
where $C_{LL}$ and $C_{LR}$ describe the contributions of new physics. The
next four terms in Eq. (\ref{e1}) with
coefficients $C_{LRLR},~C_{RLLR},~C_{LRRL}$ and $C_{RLRL}$ represent the
scalar type interactions. The remaining last two terms leaded by the
coefficients $C_T$ and $C_{TE}$ are the tensor type interactions. 

The amplitude of the exclusive $\Lambda_b \rar \Lambda\ell^+ \ell^-$ decay 
can be obtained by sandwiching ${\cal H}_{eff}$ for the $b
\rar s \ell^+ \ell^-$ transition between initial and final
baryon states, i.e., $\lla \Lambda \vel {\cal H}_{eff} \ver \lambda_b \rra$.
It follows from Eq. (\ref{e1}) that in order
to calculate the $\Lambda_b \rar \Lambda\ell^+ \ell^-$ decay amplitude
the following matrix elements are needed
\bea
&&\lla \Lambda \vel \bar s \gamma_\mu (1 \mp \gamma_5) b \ver \Lambda_b
\rra~,\nnb \\
&&\lla \Lambda \vel \bar s \sigma_{\mu\nu} (1 \mp \gamma_5) b \ver \Lambda_b
\rra~,\nnb \\
&&\lla \Lambda \vel \bar s (1 \mp \gamma_5) b \ver \Lambda_b \rra~.\nnb
\eea
Explicit forms of these matrix elements in terms of the form factors are
presented in \cite{R46212} (see also \cite{R4629}). The matrix element
of the $\Lambda_b \rar \Lambda\ell^+ \ell^-$ can be written as
\bea
\label{e3}
\lefteqn{
{\cal M} = \frac{G \alpha}{4 \sqrt{2}\pi} V_{tb}V_{ts}^\ast \Bigg\{   
\bar \ell \gamma^\mu \ell \, \bar u_\Lambda \Big[ A_1 \gamma_\mu
(1+\gamma_5) +
B_1 \gamma_\mu (1-\gamma_5) }\nnb \\
\ar i \sigma_{\mu\nu} q^\nu \big[ A_2 (1+\gamma_5) + B_2 (1-\gamma_5) \big]
+q_\mu \big[ A_3 (1+\gamma_5) + B_3 (1-\gamma_5) \big]\Big] u_{\Lambda_b}
\nnb \\
\ar \bar \ell \gamma^\mu \gamma_5 \ell \, \bar u_\Lambda \Big[
D_1 \gamma_\mu (1+\gamma_5) + E_1 \gamma_\mu (1-\gamma_5) +
i \sigma_{\mu\nu} q^\nu \big[ D_2 (1+\gamma_5) + E_2 (1-\gamma_5) \big]
\nnb \\                                                                
\ar q_\mu \big[ D_3 (1+\gamma_5) + E_3 (1-\gamma_5) \big]\Big]
u_{\Lambda_b}+
\bar \ell \ell\, \bar u_\Lambda \big(N_1 + H_1 \gamma_5\big) u_{\Lambda_b}  
+\bar \ell \gamma_5 \ell \, \bar u_\Lambda \big(N_2 + H_2 \gamma_5\big)
u_{\Lambda_b}\nnb \\
\ar 4 C_T \bar \ell \sigma^{\mu\nu}\ell \, \bar u_\Lambda \Big[ f_T    
\sigma_{\mu\nu} - i f_T^V \big( q_\nu \gamma_\mu - q_\mu \gamma_\nu \big) -
i f_T^S \big( P_\mu q_\nu - P_\nu q_\mu \big) \Big] u_{\Lambda_b}\nnb \\
\ar 4 C_{TE} \epsilon^{\mu\nu\alpha\beta} \bar \ell \sigma_{\alpha\beta}   
\ell \, i \bar u_\Lambda \Big[ f_T \sigma_{\mu\nu} -                    
i f_T^V \big( q_\nu \gamma_\mu - q_\mu \gamma_\nu \big) -
i f_T^S \big( P_\mu q_\nu - P_\nu q_\mu \big) \Big] u_{\Lambda_b}\Bigg\}~,
\eea
where $P=p_{\Lambda_b}+ p_\Lambda$.
Explicit expressions of the functions $A_i,~B_i,~D_i,~E_i,~H_j$ and $N_j$
$(i=1,2,3$ and $j=1,2)$ are given in \cite{R46212}.

From the expressions of the above-mentioned matrix elements we observe
that $\Lambda_b \rar\Lambda \ell^+\ell^-$ decay is described in terms of  
many form factors. As has already been noted, when HQET is applied the
number of independent form factors reduces to two ($F_1$ and
$F_2$) irrelevant with the Dirac structure
of the corresponding operators and it is obtained in \cite{R46210} that 
\bea
\label{e4}
\lla \Lambda(p_\Lambda) \vel \bar s \Gamma b \ver \Lambda(p_{\Lambda_b})
\rra = \bar u_\Lambda \Big[F_1(q^2) + \not\!v F_2(q^2)\Big] \Gamma
u_{\Lambda_b}~,
\eea
where $\Gamma$ is an arbitrary Dirac structure,
$v^\mu=p_{\Lambda_b}^\mu/m_{\Lambda_b}$ is the four--velocity of
$\Lambda_b$, and $q=p_{\Lambda_b}-p_\Lambda$ is the momentum transfer.
Comparing the general form of the form factors with (\ref{e5}), one can
easily obtain the following relations among them (see also \cite{R4629})
\bea
\label{e5}
g_1 \es f_1 = f_2^T= g_2^T = F_1 + \sqrt{r} F_2~, \nnb \\
g_2 \es f_2 = g_3 = f_3 = g_T^V = f_T^V = \frac{F_2}{m_{\Lambda_b}}~,\nnb \\
g_T^S \es f_T^S = 0 ~,\nnb \\
g_1^T \es f_1^T = \frac{F_2}{m_{\Lambda_b}} q^2~,\nnb \\
g_3^T \es \frac{F_2}{m_{\Lambda_b}} \ga m_{\Lambda_b} + m_\Lambda \dr~,\nnb \\
f_3^T \es - \frac{F_2}{m_{\Lambda_b}} \ga m_{\Lambda_b} - m_\Lambda \dr~,
\eea
where $r=m_\Lambda^2/m_{\Lambda_b}^2$.

Having obtained the matrix element for the $\Lambda_b \rar\Lambda \ell^+ 
\ell^-$ decay, our next aim is the calculation of lepton polarizations
with the help of this matrix element. To achieve this goal
we write the $\ell^-$ spin four--vector in terms of a unit vector
$\vec{\xi}_\mp$ along the $\ell^\mp$ spin in its rest frame as
\bea
\label{e6} 
s_\mu^\mp = \ga \frac{\vec{p}^{\;\mp}\cdot \vec{\xi}^{\;\mp}}{m_\ell},
\vec{\xi}^{\;\mp} + \frac{\vec{p}^{\;\mp} (\vec{p}^{\;\mp} \cdot
\vec{\xi}^{\;\mp})}{E_\ell+m_\ell} \dr ~,
\eea   
and choose the unit vectors along the longitudinal, normal and transversal
components of the $\ell^-$ polarization to be
\bea
\label{e7}
\vec{e}_L^{\;\mp} = \frac{\vec{p}^{\;\mp}}{\vel \vec{p}^- \ver}~, ~~~
\vec{e}_N^{\;\mp} = \frac{\vec{p}_\Lambda\times \vec{p}^{\;\mp}}
{\vel \vec{p}_\Lambda\times \vec{p}^- \ver}~,~~~
\vec{e}_T^{\;\mp} = \vec{e}_N^{\;\mp} \times \vec{e}_L^{\;\mp}~,
\eea
respectively, where  $\vec{p}^{\;\mp}$ and $\vec{p}_\Lambda$ are the three
momenta of $\ell^\mp$ and $\Lambda$, in the center of mass frame of the 
$\ell^+ \ell^-$ system. Obviously, $\vec{p}^{\;+}=-\vec{p}^{\;-}$ in this 
reference frame.

The differential decay rate of the $\Lambda_b \rar \Lambda \ell^+ \ell^-$ decay 
for any spin direction $\vec{\xi}^{\;\mp}$ along $\ell^\mp$
can be written as
\bea
\label{e8}
\frac{d\Gamma(\vec{\xi}^{\;\mp})}{ds} = \frac{1}{2}
\ga \frac{d\Gamma}{ds}\dr_0
\Bigg[ 1 + \Bigg( P_L^\mp \vec{e}_L^{\;\mp} + P_N^\mp
\vec{e}_N^{\;\mp} + P_T^\mp \vec{e}_T^{\;\mp} \Bigg) \cdot
\vec{\xi}^{\;\mp} \Bigg]~,
\eea
where $\ga d\Gamma/ds \dr_0$ corresponds to the unpolarized differential
decay rate, $s=q^2/m_{\Lambda_b}^2$ and    
$P_L^\mp$, $P_N^\mp$ and $P_T^\mp$ represent the longitudinal, normal and 
transversal polarizations of $\ell^\mp$, respectively. 
The unpolarized decay width  in Eq. (\ref{e8}) can be written as

\bea
\label{e9}
\ga \frac{d \Gamma}{s}\dr_0 = \frac{G^2 \alpha^2}{8192 \pi^5}
\vel V_{tb} V_{ts}^\ast \ver^2 \lambda^{1/2}(1,r,s) v 
\Big[{\cal T}_0(s) +\frac{1}{3} {\cal T}_2(s) \Big]~, 
\eea
where 
$\lambda(1,r,s) = 1 + r^2 + s^2 - 2 r - 2 s - 2 rs$
is the triangle function and $v=\sqrt{1-4m_\ell^2/q^2}$ is the lepton
velocity. The explicit expressions for ${\cal T}_0$ and ${\cal T}_2$ can be 
found in \cite{R46212}.

The polarizations $P_L$, $P_N$ and $P_T$ are defined as:
\bea
P_i^{(\mp)}(q^2) = \frac{\ds{\frac{d \Gamma}{ds}
                   (\vec{\xi}^{\;\mp}=\vec{e}_i^{\;\mp}) -
                   \frac{d \Gamma}{ds}
                   (\vec{\xi}^{\;\mp}=-\vec{e}_i^{\;\mp})}}
              {\ds{\frac{d \Gamma}{ds}
                   (\vec{\xi}^{\;\mp}=\vec{e}_i^{\;\mp}) +
                  \frac{d \Gamma}{ds}
                  (\vec{\xi}^{\;\mp}=-\vec{e}_i^{\;\mp})}}~, \nnb
\eea
where $i = L,N,T$. $P_L$ and $P_T$ are $P$--odd, $T$--even, while
$P_N$ is $P$--even, $T$--odd and $CP$--odd.
The explicit forms of the lepton polarizations $P_L,~P_T$ and $P_N$ can be
found in Appendix--A. 

It follows from Eq. (\ref{e11}) that the difference between $P_L^-(P_N^-)$ and  
$P_L^+(P_N^+)$ can be attributed to the existence of the scalar and tensor
interactions. Again in the same way, in massless lepton case, the difference 
between $P_T^-$ and $P_T^+$ results from scalar and vector interactions. For
this reason these above--mentioned polarizations are sensitive to the chiral
structure of the electroweak interactions and can serve as a useful tool in
search of new physics beyond the SM. 

Combined analysis of the lepton and antilepton polarizations can give
additional information about the existence of new physics, since in the SM 
$P_L^-+P_L^+=0,~P_N^-+P_N^+=0$ and $P_T^--P_T^+\simeq 0$ (in $m_\ell \rar 0$
limit). Therefore if nonzero values for the above mentioned combined
asymmetries are measured in the experiments, it can be considered as an
unambiguous indication of the existence of new physics.  

\section{Numerical analysis}

In this section we will study the dependence of the lepton polarizations, as
well as combined lepton polarization to the new Wilson coefficients. The
main input parameters in the calculations are the form factors. Since the
literature lacks exact calculations for the form factors of the $\Lambda_b
\rar \Lambda$ transition, we will use the results from QCD sum rules
approach in combination with HQET \cite{R46210,R46214}, which reduces the number of
quite many form factors into two. The $s$ dependence of these form factors
can be represented in the following way
\bea
F(q^2) = \frac{F(0)}{\ds 1-a_F s + b_F s^2}~, \nnb
\eea
where parameters $F_i(0),~a$ and $b$ are listed in table 1.
\begin{table}[h]    
\renewcommand{\arraystretch}{1.5} 
\addtolength{\arraycolsep}{3pt}  
$$
\begin{array}{|l|ccc|}
\hline
& F(0) & a_F & b_F \\ \hline
F_1 &
\phantom{-}0.462 & -0.0182 & -0.000176 \\
F_2 &
-0.077 & -0.0685 &\phantom{-}0.00146 \\ \hline
\end{array}
$$
\caption{Transition form factors for $\Lambda_b \rar \Lambda \ell^+ \ell^-$
decay in the QCD sum rules method.}
\renewcommand{\arraystretch}{1}
\addtolength{\arraycolsep}{-3pt}
\end{table}

We use the next--to--leading order logarithmic approximation for the resulting
values of the Wilson coefficients $C_9^{eff},~C_7$ and $C_{10}$ in the SM
\cite{R46215,R46216} at the renormalization point $\mu=m_b$. It should be
noted that, in addition to short distance short distance contribution,
$C_9^{eff}$ receives also long distance contributions from the real $\bar c
c$ resonant states of the $J/\psi$ family. In the present work we do not take
into account the long distance effects. In order to perform quantitative
analysis of the lepton polarizations the values of the new Wilson
coefficients, which describe the new physics beyond the SM, are needed. In
the foregoing numerical analysis we vary all new Wilson coefficients in the
range $-\vel C_{10} \ver \le C_X \le \vel C_{10} \ver$. The experimental
bounds on the branching ratio of the $B \rar K^\ast \mu^+ \mu^-$ and $B_s
\rar \mu^+ \mu^-$ \cite{R46217} suggest that this is the right order of 
magnitude range for the vector and scalar interaction coefficients.
Furthermore we assume that all new Wilson coefficients are real. 

Before performing numerical analysis, few words about lepton polarizations
are in order. From explicit expressions of the lepton polarizations one can
easily see that they depend on both $s$ and the new Wilson coefficients. For
this reason it may experimentally be problematic to study their dependence
on these variables simultaneously. Therefore we will eliminate the dependence
of the lepton polarization on one of the variables. We choose to eliminate
the variable $s$ by performing integration over $s$ in the allowed
kinematical region, so that lepton polarizations are averaged over. The
averaged lepton polarizations are defined as
\bea
\lla P_i \rra = \frac{\ds \int_{4 m_\ell^2/m_{\Lambda_b}^2}^{(1-\sqrt{r})^2}
P_i \frac{d{\cal B}}{ds} ds} 
{\ds \int_{4 m_\ell^2/m_{\Lambda_b}^2}^{(1-\sqrt{r})^2}
 \frac{d{\cal B}}{ds} ds}~.
\eea

The dependence of the averaged lepton polarizations $\lla P_L^-\rra,
~\lla P_T^-\rra$ and $\lla P_N^-\rra$ on the new Wilson coefficients are 
shown in Figs (1)--(3). From these figures we obtain the following results.

\begin{itemize}
\item $\lla P_L^-\rra$ is strongly dependent to the tensor interaction for
both $\mu$ and $\tau$ channels. In the $\tau$ channel it is also sensitive to
the scalar interaction with coefficient $C_{LRRL}$. More over it is observed
that $\lla P_L^-\rra$ is negative for all values of the new Wilson
coefficients in the $\mu$ channel, while it is positive for $C_T \simlt
-1.7$ and $C_T \simgt 0.5$ for the $\tau$ case.   

\item $\lla P_T^-\rra$ is strongly dependent to the scalar interaction 
$C_{LRRL}$, as well as to the tensor interaction for the $\mu$ channel.
The $\tau$ channel is strongly dependent on the tensor interaction. From
Fig. (2a) we see that $\lla P_T^-\rra$ is negative (positive) for the $\mu$
channel when $C_{LRRL} \simlt -1.5~(C_{LRRL} \simgt -1.5),~C_T \simgt 
0.75~(C_T \simlt 0.75)$ and $C_{TE} \simlt -0.5~(C_{LRRL} \simgt -0.5)$.
For $\tau$ channel the situation is different and $\lla P_T^-\rra$ is
positive when $C_T \simlt -2.6$ and $C_T \simgt 0.6$ and is negative for all
other values of the new Wilson coefficients. 

\item It follows from Eq. (\ref{e13}) that the normal polarization is proportional
to the imaginary parts of the combination of the products of the new Wilson 
coefficients. However, since in this work we assume that all new Wilson
coefficients are real, the nonzero value of $\lla P_N^-\rra$ is due to the
imaginary part of the $C_9^{eff}$ only. Moreover, since $\lla P_N^-\rra$ is
proportional to the lepton mass, it is maximum value is around $1\%$ for the $\mu$
channel and therefore we do not present its dependence on the new Wilson
coefficients. For the $\tau$ case $\lla P_N^-\rra$ shows stronger dependence
on $C_{LL},~C_{LRRL}$, as well as on $C_T$. It is interesting to note that
$\lla P_N^-\rra$ is positive when $C_T \simlt -1.25,~c_T \simgt 0.7$, while
it is negative for all other cases.   
\end{itemize}

Our numerical analysis for the combined lepton, antilepton polarizations
leads to the following results.

\begin{itemize}
\item $\lla P_L^- + P_L^+ \rra$ is strongly dependent only to the scalar
type interactions and quite weakly on the remaining new Wilson coefficients.
$\lla P_L^- + P_L^+ \rra$ is positive (negative) when $C_{RLRL}$,
$C_{LRRL}$ are negative (positive) and $C_{RLLR}$, $C_{LRLR}$ are positive
(negative). The magnitude of $\lla P_L^- + P_L^+ \rra$ for the $\mu$ channel
varies between $-0.15$ and $0.15$ depending on the variation of the
scalar interaction. 

For $\tau$ case, $\lla P_L^- + P_L^+ \rra$ exhibits strong dependence on 
the tensor interaction $C_T$ and scalar interactions $C_{LRLR}$ and
$C_{LRRL}$. $\lla P_L^- + P_L^+ \rra$ is positive (negative) when $C_T
\simlt 0~(C_T\simgt 0)$ and the sign of the scalar interactions are negative
(positive). The value of $\lla P_L^- + P_L^+ \rra$ for this case varies
between $-0.35$ and $0.4$ depending on the variation of the
corresponding tensor and scalar interactions. It should be noted that 
$\lla P_L^- + P_L^+ \rra=0$ in the SM.

\item The situation for the combined $\lla P_T^- - P_T^+ \rra$ 
polarization is as follows. For the $\mu$ channel, $\lla P_T^- - P_T^+ \rra$ 
is strongly dependent on the tensor interactions $C_T$, $C_{TE}$ and the 
scalar $C_{LRLR},~C_{LRRL}$ coefficients when $C_{TE}\simlt -0.5$, $C_T\simgt 0.9$,
$C_{LRLR}\simgt 2$ and $C_{LRRL}\simlt -2$, $\lla P_T^- - P_T^+ \rra$ is
negative and for all other cases it is positive. The magnitude of 
$\lla P_T^- - P_T^+ \rra$ varies between the values $(-0.2 \div 0.45)$
depending on the variation of the scalar and tensor interaction
coefficients.

When we investigate the
$\tau$ case, $\lla P_T^- - P_T^+ \rra$ is observed to be strongly dependent on
tensor interactions. $\lla P_T^- - P_T^+ \rra$  is negative only 
for $C_{TE} \simgt 1.2$, otherwise it is positive. The magnitude of $\lla
P_T^- - P_T^+ \rra$ for the $\tau$ channel lies in the region $(-0.01 \div
0.98)$ depending on the variation of the tensor interaction coefficient.

\item The final analysis we present is the combined $\lla P_N^- +
P_N^+ \rra$ polarization, which essentially shows dependence only on $C_T$
and $C_{TE}$. Its value ranges between $-0.015$ and $0.035$ depending on the
variation of the tensor interactions.   
\end{itemize}

From these analyzes we can conclude that the change in sign and magnitude of
both $\lla P_i^- \rra$ and $\lla P_i^- +(-) P_i^+\rra$ (-- sign is for the
transversal polarization case) is an indication of the existence of new
physics beyond the SM.

At the and of our analysis, we would like to discuss about the following
problem. The branching ratio of the $\Lambda_b \rar \Lambda \ell^- \ell^+$
decay depends also on the new Wilson coefficients. Moreover its experimental
measurement is easier compared to the case with the lepton polarizations.
One could ask then what advantage the measurement with lepton polarizations
has, since similar information can be attained from the branching ratio
measurement. Obviously, measurement of lepton polarizations are quite useful
for establishing new physics if we can find the parameter space of the new 
Wilson coefficients in which the decay rate agrees with the SM while the
lepton polarizations deviate considerably from the SM case. The intriguing
question is whether there exist such a region for the new Wilson
coefficients $C_X$. In order to answer this question we present in Figs.
(4)--(8) the correlation between the branching ratio and the averaged
and averaged--combined lepton polarizations for the $\Lambda_b \rar \Lambda
\ell^- \ell^+$ decay. In Fig. (4) we present the correlation in the 
$\ga {\cal B},\lla P_L^- \rra \dr$ plane by varying the coefficients of the
scalar, vector and tensor type interactions. Depicted in Fig. (5) is the
correlation in the $\ga {\cal B},\lla P_L^- + P_L^+\rra \dr$ plane by varying the
coefficients of the new scalar, vector and tensor type interactions. In
these figures the value of the branching ratio is restricted to have the
values in the region $2\times 10^{-7} \le {\cal B}(\Lambda_b \rar \Lambda \tau^+
\tau^- ) \le 5\times 10^{-7}$ which is quite close to the SM prediction. 

We observe from these figures that, indeed there exist regions for the new
Wilson coefficients for which $\lla P_L^- \rra$ and $\lla P_L^-+P_L^+ \rra$
show considerable departure from the SM prediction, while branching ratio
coincide with the SM result. 

Our numerical results show that the case for the transversal, normal
and the combined polarizations $\lla P_T^--P_T^+ \rra$ and $\lla P_N^-+P_N^+
\rra$, one can find similar regions of the new Wilson coefficients with
branching ratios coinciding with the SM results while lepton polarizations
differing considerably (see Figs. (7),(8)).

In conclusion, we present the most general analysis of the lepton
polarizations in the exclusive $\Lambda_b \rar \Lambda \ell^- \ell^+$ decay,
using the general, model independent form of the effective Hamiltonian. The
sensitivity of the longitudinal, transversal and normal polarizations of
$\ell^-$, as well as lepton--antilepton combined asymmetries on the new
Wilson coefficients, are studied. We find out that there exist regions in
the parameter space of the new Wilson coefficients, in which branching
ratios coincide with the SM results while lepton polarizations  
differ considerably. A thorough study of the lepton polarizations in this
region of the parameter space of the new Wilson coefficients can establish
new physics beyond the SM.

\newpage

\bAPP{A}{Lepton polarizations}

In this appendix we present the explicit form of the expressions for the
longitudinal $P_L$, transversal $P_T$ and normal $P_N$ 
lepton polarizations. The --(+) sign in these formulas corresponds to the
particle (antiparticle), respectively.

\baeeq
\label{e11}
P_L^\mp \es \frac{256}{3} \lambda m_\ell m_{\Lambda_b}^5 v (1+\sqrt{r})
\, \mbox{\rm Re}[(A_1+B_1)^\ast C_{TE} f_T^S\pm  
(D_1+E_1)^\ast C_T f_T^S] \nnb \\
\ar 32 m_\ell m_{\Lambda_b}^3 v (1+\sqrt{r}) [s-(1-\sqrt{r})^2] \,   
\mbox{\rm Re}[(D_1-E_1)^\ast H_1] \nnb \\
\ek 32 m_\ell m_{\Lambda_b}^3 v (1- \sqrt{r}) [s-(1+\sqrt{r})^2] \,   
\mbox{\rm Re}[(D_1+E_1)^\ast F_1] \nnb \\
\ar \frac{512}{3} \lambda m_{\Lambda_b}^6 s v \Big(
8 \, \mbox{\rm Re}[C_T^\ast C_{TE}] \, \mbox{\rm Re}[f_T^\ast f_T^S] -
m_\ell \, 
\mbox{\rm Re}[(A_2+B_2)^\ast C_{TE} f_T^S \pm
(D_2+E_2)^\ast C_T f_T^S] \Big) \nnb \\
\ek \frac{4096}{3} \lambda m_{\Lambda_b}^7 s v (1+\sqrt{r})\,
\mbox{\rm Re}[C_T^\ast C_{TE}] \, \mbox{\rm Re}[f_T^{S\ast} f_T^V] \nnb \\
\kpm 64 m_{\Lambda_b}^4 s v \sqrt{r} \Big(
2 \, \mbox{\rm Re}[A_1^\ast E_1 + B_1^\ast D_1] -
m_{\Lambda_b} (1-r+s) \, \mbox{\rm Re}[A_1^\ast D_2+A_2^\ast D_1]                
\Big) \nnb \\
\kmp 64 m_{\Lambda_b}^5 \sqrt{r} (1-r+s) s v \,
\mbox{\rm Re}[B_1^\ast E_2 + B_2^\ast E_1] \nnb \\
\ek 512 m_{\Lambda_b}^5 s v (1+\sqrt{r}) [s - (1-\sqrt{r})^2] \Big(
8 \, \mbox{\rm Re}[C_T^\ast C_{TE}] \, \mbox{\rm Re}[f_T^\ast f_T^V] \nnb \\
\ek m_\ell \, \mbox{\rm Re}[(A_2+B_2)^\ast C_{TE} f_T^V \pm
(D_2+E_2)^\ast C_T f_T^V] \Big) \nnb \\
\ar \frac{2048}{3} \lambda m_{\Lambda_b}^8 s v [s - (1+\sqrt{r})^2]
\vel f_T^S \ver^2 \, \mbox{\rm Re}[C_T^\ast C_{TE}] \nnb \\
\kpm 128 m_{\Lambda_b}^6 s^2 v \sqrt{r} \,
\mbox{\rm Re}[A_2^\ast E_2 + B_2^\ast D_2] \nnb \\
\ek 16 m_{\Lambda_b}^4 s v [s-(1+\sqrt{r})^2] \,
\mbox{\rm Re}[F_1^\ast F_2 + 2 m_\ell (D_3+E_3)^\ast F_1] \nnb \\
\ek 16 m_{\Lambda_b}^4 s v [s-(1-\sqrt{r})^2] \,
\mbox{\rm Re}[H_1^\ast H_2 + 2 m_\ell (D_3-E_3)^\ast H_1] \\
\kpm 64 m_{\Lambda_b}^5 s v (1-r-s) \,\mbox{\rm Re}[A_1^\ast E_2 +
A_2^\ast E_1 + B_1^\ast D_2 + B_2^\ast D_1] \nnb \\
\kmp \frac{64}{3} m_{\Lambda_b}^4 v [ 1+r^2 +r (s-2)+s(1-2 s)] \,
\mbox{\rm Re}[A_1^\ast D_1 + B_1^\ast E_1] \nnb \\
\ar \frac{512}{3} m_{\Lambda_b}^4 v [(1-r)^2 +(1-6 \sqrt{r} +r) s -2 s^2]
\Big( m_\ell \, \mbox{\rm Re}[(A_1+B_1)^\ast C_{TE} f_T^V \pm   
(D_1+E_1)^\ast C_T f_T^V] \nnb \\
\ek 4 m_{\Lambda_b}^2 s \vel f_T^V \ver^2 \,\mbox{\rm Re}[C_T^\ast C_{TE}]
\Big) \nnb \\
\ek \frac{4096}{3} m_{\Lambda_b}^4 v [2 (1-r)^2 - (1+r) s -s^2]   
\vel f_T \ver^2 \,\mbox{\rm Re}[C_T^\ast C_{TE}] \nnb \\
\ar \frac{256}{3} m_\ell m_{\Lambda_b}^4 v \Big\{
(1+2 \sqrt{r}+r-s) (2-4 \sqrt{r}+2 r +s)\, 
\mbox{\rm Re}[(B_2-A_2)^\ast C_T f_T]\nnb \\
\ar 2 (1-2 \sqrt{r}+r-s) (2+4 \sqrt{r}+2 r +s)\,
\mbox{\rm Re}[(B_2+A_2)^\ast C_{TE} f_T] \Big\} \nnb \\
\kpm \frac{512}{3} m_\ell m_{\Lambda_b}^4 v \Big\{
(1-2 \sqrt{r}+r-s) (2+4 \sqrt{r}+2 r +s) \,
\mbox{\rm Re}[(D_2+E_2)^\ast C_T f_T] \nnb \\
\ek 2(1+2 \sqrt{r}+r-s) (2-4 \sqrt{r}+2 r +s)\, 
\mbox{\rm Re}[(D_2-E_2)^\ast C_{TE} f_T] \Big\} \nnb \\
\kmp \frac{64}{3} m_{\Lambda_b}^6 s v [2 + r (2 r -4 -s) -s (1+s)] \,
\mbox{\rm Re}[A_2^\ast D_2 + B_2^\ast E_2] \nnb \\
\kmp 512 m_\ell m_{\Lambda_b}^3 v \Big\{
(1 + \sqrt{r}) (1 - 2 \sqrt{r} + r - s)\,
\mbox{\rm Re}[(D_1+E_1)^\ast C_T f_T] \nnb \\
\ar 2(1 - \sqrt{r}) (1 + 2 \sqrt{r} + r - s)\,
\mbox{\rm Re}[(D_1-E_1)^\ast C_{TE} f_T] \Big\} \nnb \\
\ar 256 m_\ell m_{\Lambda_b}^3 v \Big\{      
(1 - \sqrt{r}) (1 + 2 \sqrt{r} + r - s)\,
\mbox{\rm Re}[(B_1-A_1)^\ast C_T f_T] \nnb \\
\ek 2(1 + \sqrt{r}) (1 - 2 \sqrt{r} + r - s)\,
\mbox{\rm Re}[(B_1+A_1)^\ast C_{TE} f_T] \Big\}~,
\nnb \\ \nnb \\
\label{e12}
P_T^\mp \es -16 \pi m_\ell m_{\Lambda_b}^3 \sqrt{s\lambda} 
\Big( \vel A_1 \ver^2  - \vel B_1 \ver^2 \Big) \nnb \\
\kmp 64 \pi m_\ell m_{\Lambda_b}^3 \sqrt{s\lambda} \,
\mbox{\rm Re}[(C_T F_2^\ast - 2 C_{TE} H_2^\ast) f_T] \nnb \\
\kpm 128 \pi m_\ell^2 m_{\Lambda_b}^3 \sqrt{s\lambda} \,
\mbox{\rm Re}[2 (D_3-E_3)^\ast C_{TE} f_T - 
(D_3+E_3)^\ast C_T f_T] \nnb \\
\ar 32 \pi m_\ell m_{\Lambda_b}^4 \sqrt{s\lambda} \, \mbox{\rm Re}[A_1^\ast
B_2-A_2^\ast B_1] \nnb \\
\kmp 16 \pi m_\ell m_{\Lambda_b}^4 \sqrt{s\lambda} \, 
\mbox{\rm Re}[A_1^\ast E_3 - A_2^\ast E_1 + B_1^\ast D_3 -
B_2^\ast D_1] \nnb \\
\ar 4 \pi m_{\Lambda_b}^4 \sqrt{s\lambda} (1-\sqrt{r}) \Big(
\pm \mbox{\rm Re}[(A_1-B_1)^\ast H_2] +
512 m_\ell \, \mbox{\rm Re}[C_T C_{TE}^\ast f_T^\ast f_T^V] \Big) \nnb \\
\kmp 4 \pi m_{\Lambda_b}^4 \sqrt{s\lambda} (1+\sqrt{r}) \Big(
\mbox{\rm Re}[(A_1+B_1)^\ast F_2] - 16 m_\ell \, \mbox{\rm Re}[C_T F_2^\ast
f_T^V] \nnb \\
\ek 32 m_\ell^2 \, \mbox{\rm Re}[(D_3+E_3)^\ast C_T f_T^V] \Big)\nnb \\
\ar 16 \pi m_\ell m_{\Lambda_b}^5 \sqrt{s\lambda} (1-r) \Big(                 
\vel A_2 \ver^2 - \vel B_2 \ver^2 \Big) \nnb \\
\ar 16 \pi m_\ell m_{\Lambda_b}^4 \sqrt{rs\lambda} \, \mbox{\rm Re}
[2 A_1^\ast A_2 -  2 B_1^\ast B_2 \mp A_1^\ast D_3 \mp 
A_2^\ast D_1 \mp B_1^\ast E_3 \mp B_2^\ast E_1] \nnb \\
\ek 16\pi m_\ell m_{\Lambda_b}^3 \sqrt{\frac{\lambda}{s}}(1-r)
\Big( \pm \mbox{\rm Re}[A_1^\ast D_1 + B_1^\ast E_1] +
128 \vel f_T \ver^2 \, \mbox{\rm Re}[C_T^\ast C_{TE}] \Big) \\
\kpm 4 \pi m_{\Lambda_b}^5 s \sqrt{s\lambda} \Big(
\mbox{\rm Re}[(A_2+B_2)^\ast F_2 + (A_2-B_2)^\ast H_2] +
4 m_\ell \, \mbox{\rm Re}[A_2^\ast D_3 + B_2^\ast E_3] \Big) \nnb \\
\kpm 128 \pi m_\ell^2 m_{\Lambda_b}^4 \sqrt{\frac{\lambda}{s}}
(1-\sqrt{r}) [(1+\sqrt{r})^2 -s] \, 
\mbox{\rm Re}[(D_1+E_1)^\ast C_T f_T^S] \nnb \\
\kpm 64 \pi m_\ell m_{\Lambda_b}^5 \sqrt{s\lambda} [(1+\sqrt{r})^2 -s]
\Big( \mbox{\rm Re}[F_2^\ast C_T f_T^S] +
2 m_\ell \, \mbox{\rm Re}[(D_3+E_3)^\ast C_T f_T^S] \Big) \nnb \\
\kmp 32 \pi m_{\Lambda_b}^4 \sqrt{s\lambda} (1-2 v^2) 
\Big\{ (1-\sqrt{r}) \, \mbox{\rm Re}[(D_1+E_1)^\ast C_T f_T] \nnb \\
\ar 2 (1+\sqrt{r}) \, \mbox{\rm Re}[(D_1-E_1)^\ast C_{TE} f_T] \Big\} \nnb \\
\ek 32 \pi m_{\Lambda_b}^4 \sqrt{s\lambda} (2-v^2)
\Big\{ (1+\sqrt{r}) \, \mbox{\rm Re}[(A_1-B_1)^\ast C_T f_T] \nnb \\ 
\ar 2 (1-\sqrt{r}) \, \mbox{\rm Re}[(A_1+B_1)^\ast C_{TE} f_T] \Big\} \nnb \\
\ar 32 \pi m_{\Lambda_b}^5 s \sqrt{s\lambda} (2-v^2)
\Big( \mbox{\rm Re}[(A_1-B_1)^\ast C_T f_T^V] +
m_{\Lambda_b} (1-\sqrt{r}) \, 
\mbox{\rm Re}[(A_2-B_2)^\ast C_T f_T^V] \Big)\nnb \\
\ek 32 \pi m_{\Lambda_b}^5 \sqrt{s\lambda} (1-r) (2-v^2)
\Big( \mbox{\rm Re}[(A_2-B_2)^\ast C_T f_T] -
2 \, \mbox{\rm Re}[(A_2+B_2)^\ast C_{TE} f_T] \Big) \nnb \\
\ar 4 \pi m_{\Lambda_b}^4 \sqrt{s\lambda} v^2 
\Big\{ (1-\sqrt{r}) \, \mbox{\rm Re}[(D_1-E_1)^\ast H_1] -
(1+\sqrt{r}) \, \mbox{\rm Re}[(D_1+E_1)^\ast F_1] \Big\} \nnb \\
\kmp 32 \pi m_{\Lambda_b}^5 \sqrt{s\lambda} (1-r) v^2 \Big(
\mbox{\rm Re}[(D_2+E_2)^\ast C_T f_T] -   
2 \, \mbox{\rm Re}[(D_2-E_2)^\ast C_{TE} f_T] \Big)\nnb \\
\ar 4 \pi m_{\Lambda_b}^5 s \sqrt{s\lambda} v^2 \Big(
\mbox{\rm Re}[(D_2+E_2)^\ast F_1] +   
\mbox{\rm Re}[(D_2-E_2)^\ast H_1] \Big) \nnb \\
\kmp 64 \pi m_{\Lambda_b}^6 s \sqrt{s\lambda}(1-\sqrt{r}) v^2 \,     
\mbox{\rm Re}[(D_2-E_2)^\ast C_{TE} f_T^V] \nnb \\
\kpm 64 \pi m_{\Lambda_b}^3 \sqrt{\frac{\lambda}{s}} \Big\{
2 m_\ell^2 (1-r) \, \mbox{\rm Re}[(D_1+E_1)^\ast C_T f_T^V] -
m_{\Lambda_b}^2 s^2 v^2 \,
\mbox{\rm Re}[(D_1-E_1)^\ast C_{TE} f_T^V] \Big\}~, \nnb \\
\nnb \\ \nnb \\
\label{e13}
P_N^\mp \es
\pm 16  \pi m_\ell m_{\Lambda_b}^3 v \sqrt{s\lambda}
\Big( \mbox{\rm Im}[A_1^\ast D_1 - B_1^\ast E_1] +                               
4 \, \mbox{\rm Im}[F_1^\ast C_T f_T] - 8 \, \mbox{\rm Im}[H_1^\ast C_{TE} f_T]
\Big) \nnb \\
\ar 16  \pi m_\ell m_{\Lambda_b}^4 v \sqrt{s\lambda} \Big(
\pm \mbox{\rm Im}[B_1^\ast D_2 - A_1^\ast E_2] + 
\mbox{\rm Im}[(\pm A_2+D_2+D_3)^\ast E_1] \nnb \\ 
\ek \mbox{\rm Im}[(\pm B_2-E_2-E_3)^\ast D_1] \Big) \nnb \\
\ar 32 \pi m_{\Lambda_b}^4 v \sqrt{s\lambda} \Big( 
\mbox{\rm Im}[B_1^\ast (C_T-2 C_{TE}) f_T] -
\sqrt{r} \, \mbox{\rm Im}[B_1^\ast (C_T+2 C_{TE}) f_T] \Big) \nnb \\
\ar 4 \pi m_{\Lambda_b}^4 v \sqrt{s\lambda} \Big\{ (1-\sqrt{r})\,
\mbox{\rm Im}[\pm (A_1-B_1)^\ast H_1 + (D_1-E_1)^\ast H_2] \nnb \\
\ek (1+\sqrt{r}) \, \mbox{\rm Im}[\pm (A_1+B_1)^\ast F_1 + (D_1+E_1)^\ast F_2] \nnb \\ 
\ek 128 m_\ell (1-\sqrt{r}) \ga 4 \vel C_{TE} 
\ver^2 - \vel C_T \ver^2 \dr \,
\mbox{\rm Im}[f_T^\ast f_T^V] \Big\} \nnb \\
\kmp 64 \pi m_\ell m_{\Lambda_b}^4 v \sqrt{s\lambda} (1+\sqrt{r}) \,      
\mbox{\rm Im}[F_1^\ast C_T f_T^V] \nnb \\
\ar 32  \pi m_{\Lambda_b}^4 v \sqrt{s\lambda} \Big(
\mbox{\rm Im}[(A_1 \mp D_1 \pm \sqrt{r} E_1)^\ast (C_T+2 C_{TE}) f_T]  \\
\ek \mbox{\rm Im}[(\sqrt{r} A_1 \pm \sqrt{r} D_1 \mp E_1)^\ast (C_T-2 C_{TE}) f_T]
\Big) \nnb \\
\ek 32  \pi m_{\Lambda_b}^5 v \sqrt{s\lambda}(1-r) \Big(
\mbox{\rm Im}[(A_2 \pm D_2)^\ast (C_T-2 C_{TE}) f_T] +
\mbox{\rm Im}[(B_2 \mp E_2)^\ast (C_T+2 C_{TE}) f_T] \Big) \nnb \\
\kmp 16  \pi m_\ell m_{\Lambda_b}^4 v \sqrt{s\lambda} \Big(
m_{\Lambda_b} (1-r) \, \mbox{\rm Im}[A_2^\ast D_2-B_2^\ast E_2] +
\sqrt{r} \, \mbox{\rm Im}[A_1^\ast D_2 +A_2^\ast D_1] \Big) \nnb \\
\ar 16  \pi m_\ell m_{\Lambda_b}^4 v \sqrt{r s\lambda} \,       
\mbox{\rm Im}[D_1^\ast (D_2-D_3) - E_2^\ast (\pm B_1+E_1) -
E_1^\ast (\pm B_2+E_3)] \nnb \\
\ar 4 \pi m_{\Lambda_b}^5 v s \sqrt{s\lambda} \,      
\mbox{\rm Im}[\pm (A_2+B_2)^\ast F_1 + (D_2+E_2)^\ast F_2] \nnb \\
\ek 32 \pi m_{\Lambda_b}^5 v s \sqrt{s\lambda} \,      
\mbox{\rm Im}[2 (A_1-B_1)^\ast C_{TE} f_T^V \mp 
(D_1-E_1)^\ast C_T f_T^V] \nnb \\
\ar 4 \pi m_{\Lambda_b}^5 v s \sqrt{s\lambda} \Big(
\mbox{\rm Im}[\pm (A_2-B_2)^\ast H_1 + (D_2-E_2)^\ast H_2] +
4 m_\ell \, \mbox{\rm Im}[D_2^\ast D_3 + E_2^\ast E_3] \Big) \nnb \\
\ek 32 \pi m_{\Lambda_b}^6 v s \sqrt{s\lambda}(1-\sqrt{r}) \Big(
\mbox{\rm Im}[2 (A_2-B_2)^\ast C_{TE} f_T^V \mp
(D_2-E_2)^\ast C_T f_T^V] \nnb \\
\kmp 64 \pi m_\ell m_{\Lambda_b}^5 v \sqrt{s\lambda}
[(1+\sqrt{r})^2 - s] \, \mbox{\rm Im}[F_1^\ast C_T f_T^S]~,\nnb
\eaeeq

\eAPP

\newpage

\section*{Figure captions}
{\bf Fig. (1)} The dependence of the averaged longitudinal lepton
polarization $\lla P_L^- \rra$ on the new Wilson coefficients for the
$\Lambda_b \rar \Lambda \ell^+ \ell^-,~(\ell=\mu,~\tau)$ decay.\\ \\
{\bf Fig. (2)} The same as in Fig. (1), but for the averaged transversal
lepton polarization $\lla P_T^- \rra$.\\ \\
{\bf Fig. (3)} The dependence of the averaged normal lepton
polarization $\lla P_N^- \rra$ on the new Wilson coefficients for the
$\Lambda_b \rar \Lambda \tau^+ \tau^-$ decay.\\ \\
{\bf Fig. (4)} Parametric plot of the correlation between the branching
ratio ${\cal B}$ (in units of $10^{-7}$) and the averaged longitudinal 
polarization $\lla P_L^- \rra$ as a function of the new Wilson coefficients
for the $\Lambda_b \rar \Lambda \tau^+ \tau^-$ decay.\\ \\
{\bf Fig. (5)} The same as in Fig. (4), but for the combined averaged
longitudinal lepton polarization $\lla P_L^- + P_L^+ \rra$.\\ \\
{\bf Fig. (6)} The same as in Fig. (4), but for the averaged transversal
lepton polarization $\lla P_T^- \rra$. \\ \\
{\bf Fig. (7)} The same as in Fig. (4), but for the combined averaged 
transversal lepton polarization $\lla P_T^- - P_T^+\rra$. \\ \\
{\bf Fig. (8)} The same as in Fig. (4), but for the combined averaged
normal lepton polarization $\lla P_N^- + P_N^+\rra$.

\newpage

\begin{figure}
\vskip 1.5 cm
    \includegraphics{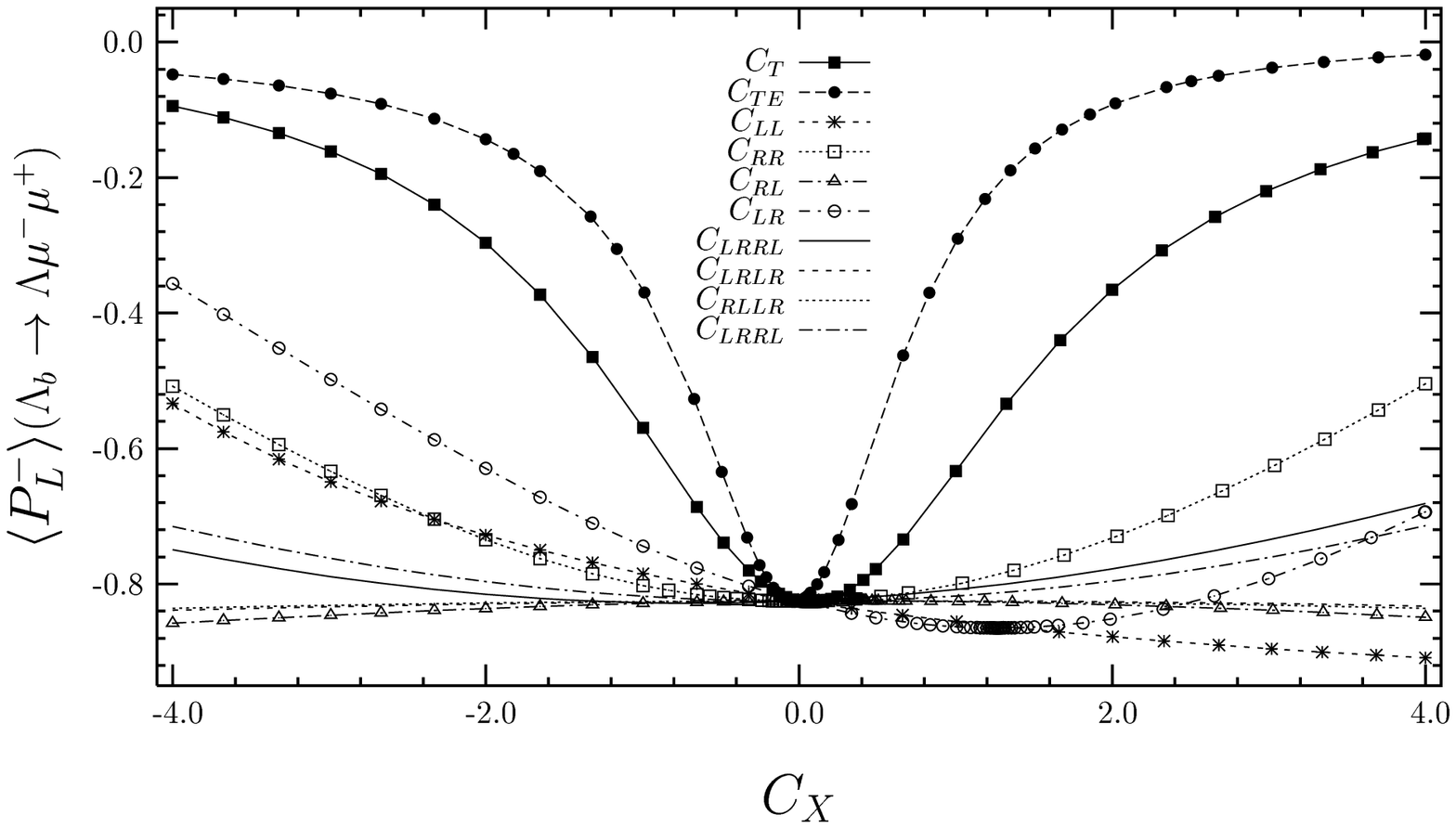}
\vskip 7.8cm
\begin{center}
{\bf Fig. 1--a} 
\end{center}
\end{figure}  

\begin{figure}   
\vskip 2.5 cm
    \includegraphics{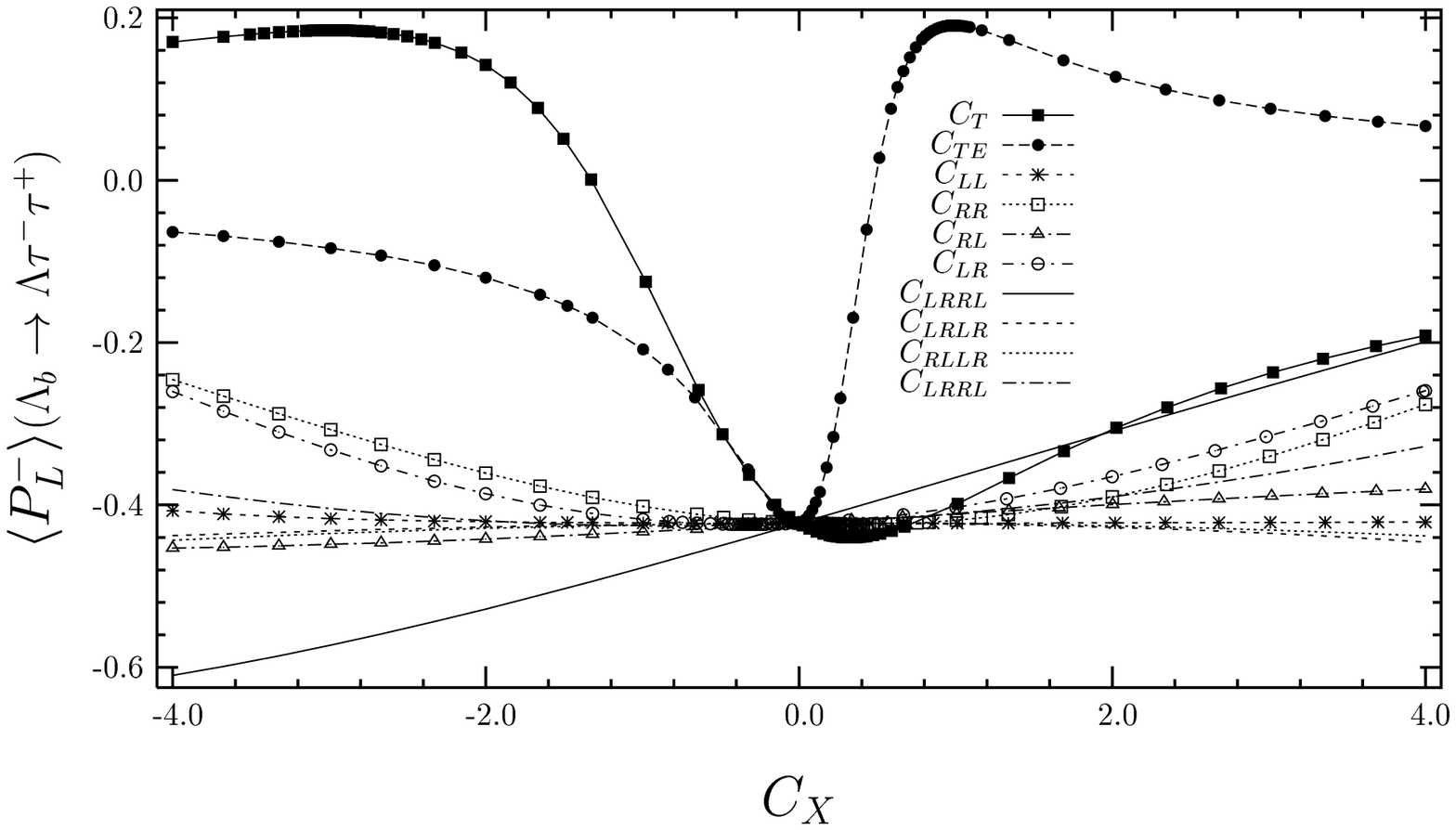}
\vskip 7.8 cm   
\begin{center}
{\bf Fig. 1--b}
\end{center}
\end{figure}

\begin{figure}   
\vskip 1.5 cm
    \includegraphics{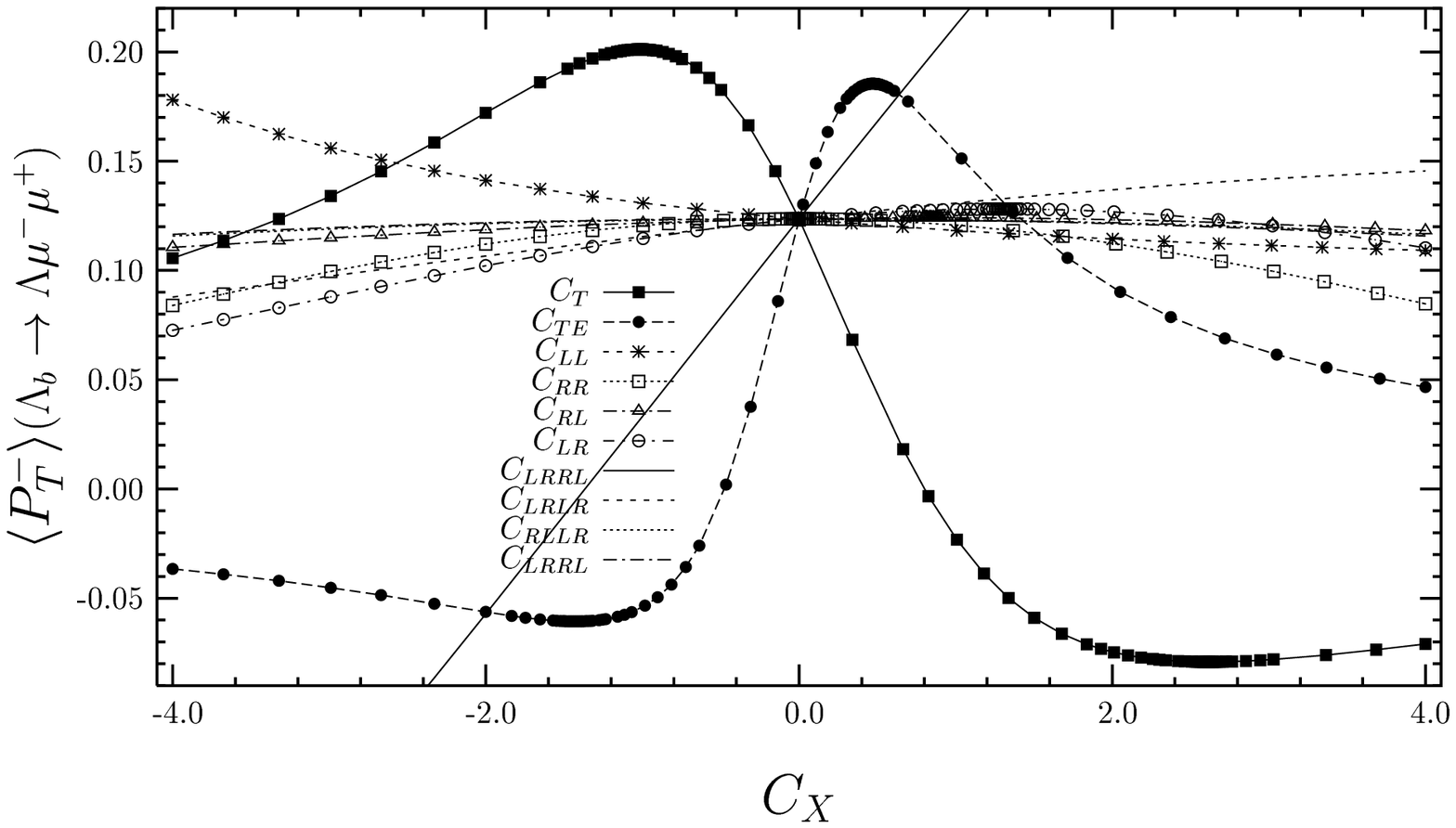}
\vskip 7.8cm
\begin{center}
{\bf Fig. 2--a}
\end{center}
\end{figure}

\begin{figure}    
\vskip 2.5 cm
    \includegraphics{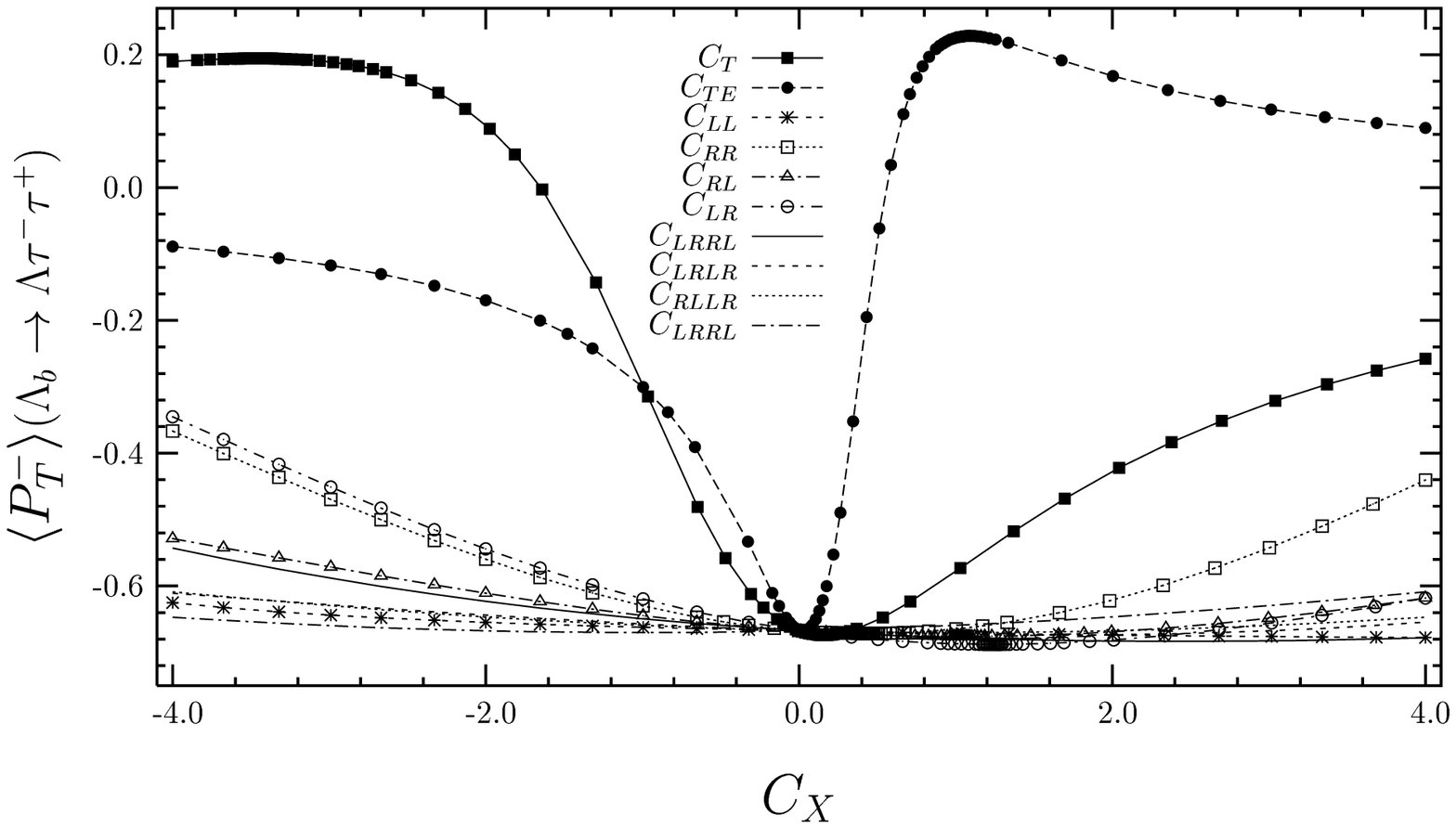}
\vskip 7.8 cm   
\begin{center}
{\bf Fig. 2--b}
\end{center}
\end{figure}

\begin{figure}
\vskip 1.5 cm
    \includegraphics{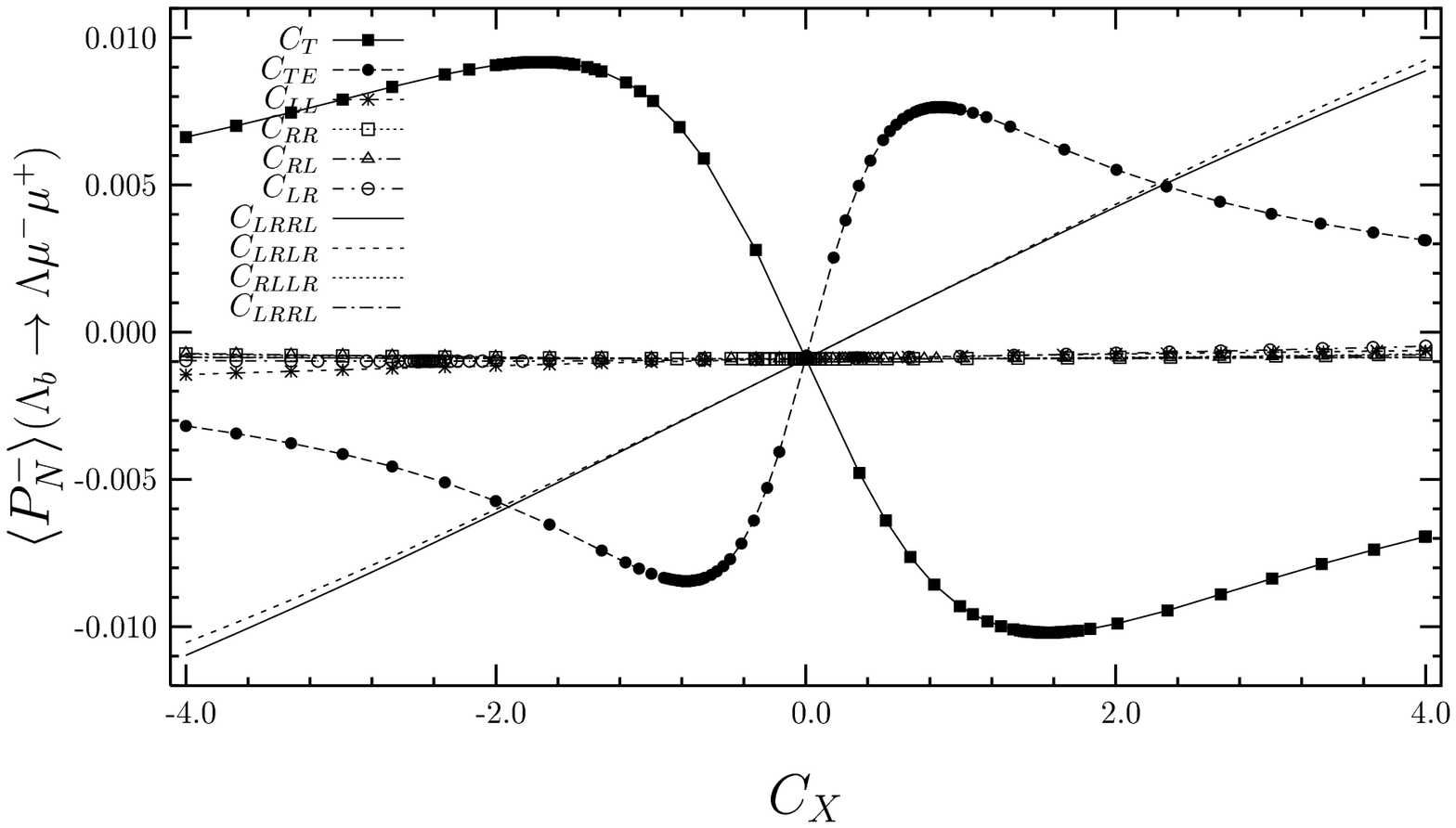}
\vskip 7.8cm
\begin{center}
{\bf Fig. 3--a}
\end{center}
\end{figure}

\begin{figure}   
\vskip 2.5 cm
    \includegraphics{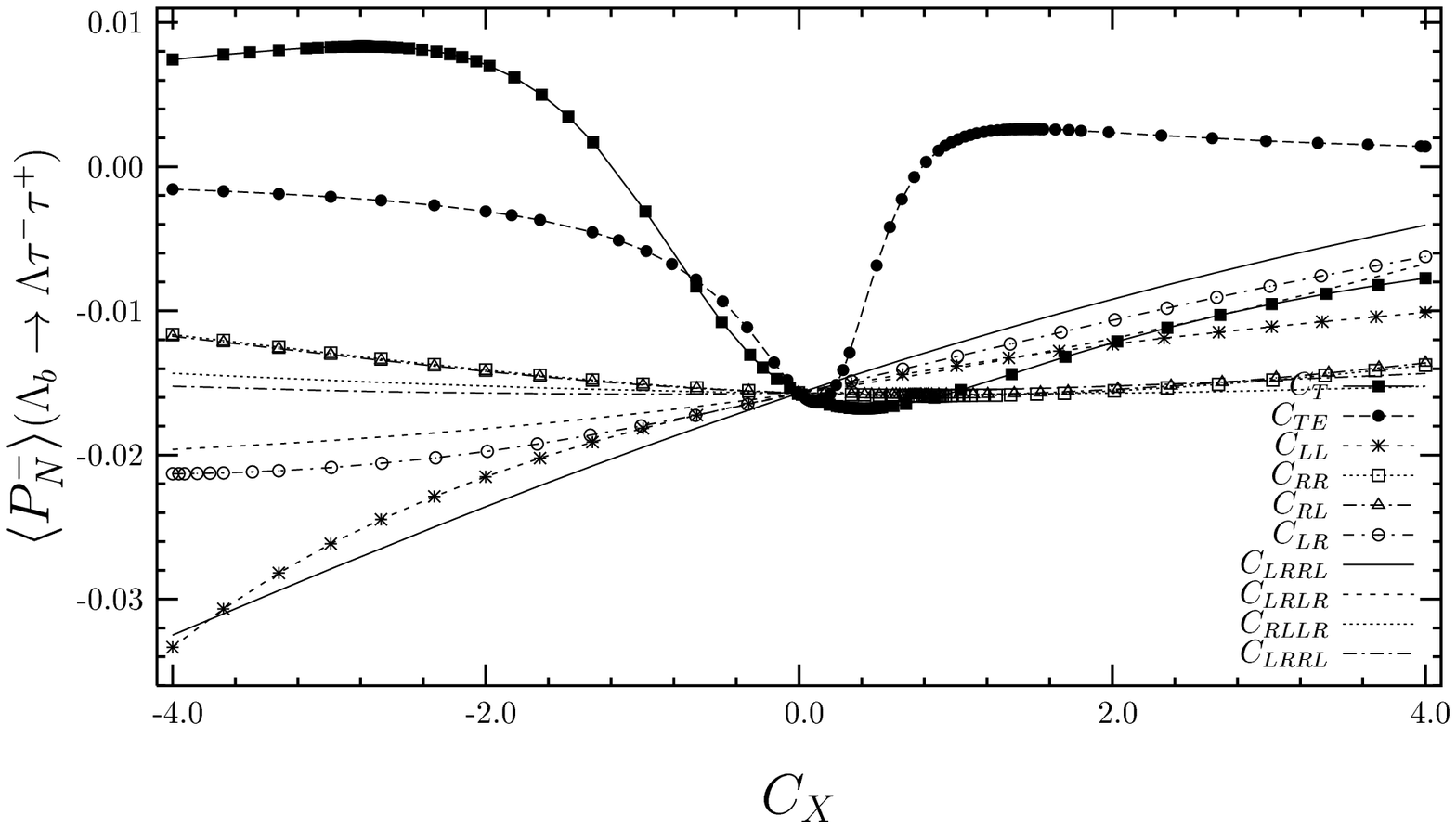}
\vskip 7.8 cm    
\begin{center}
{\bf Fig. 3--b}
\end{center}
\end{figure}

\begin{figure}
\vskip 1.5 cm
    \includegraphics{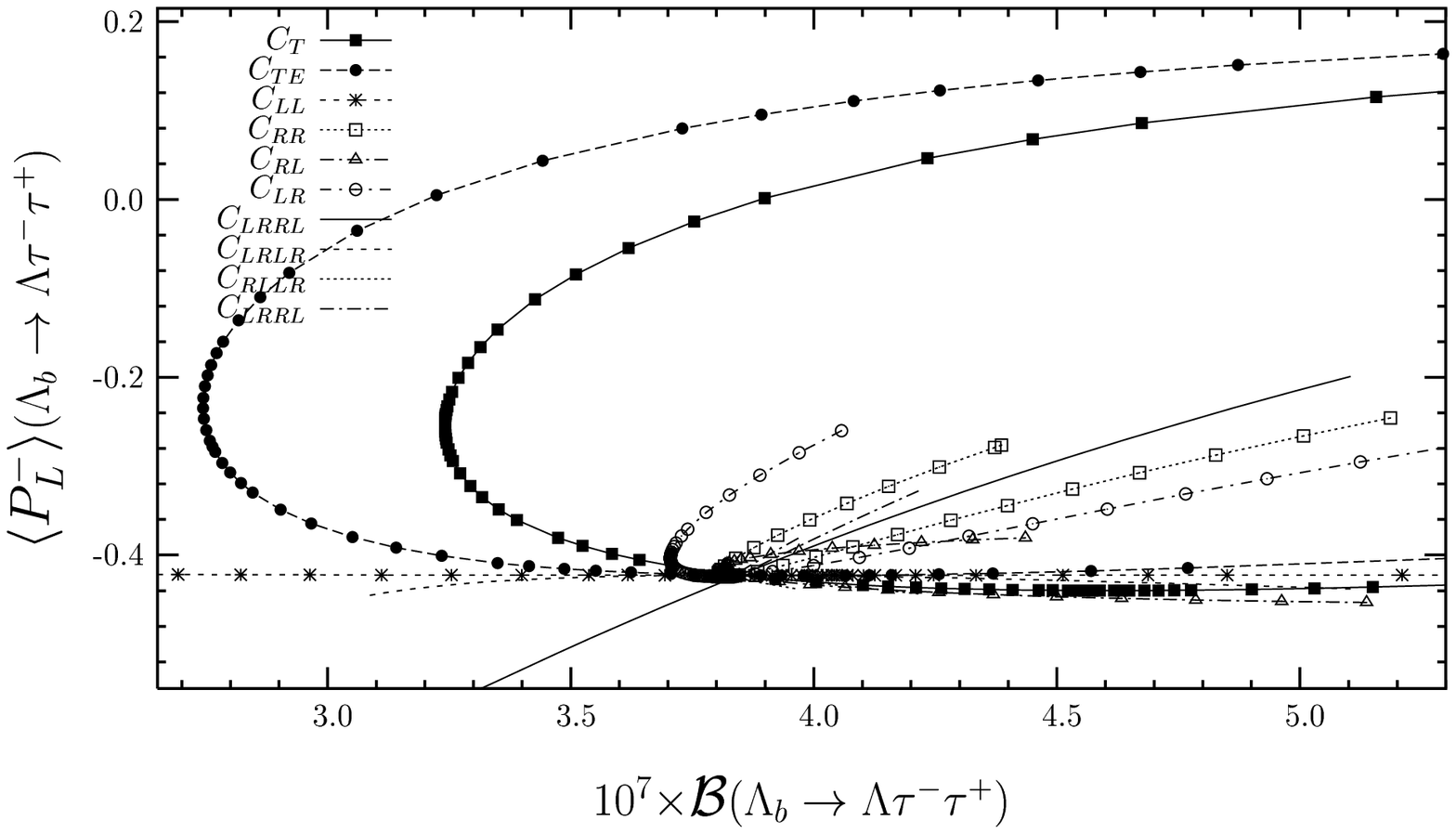}
\vskip 7.8cm
\begin{center}
{\bf Fig. 4}
\end{center}
\end{figure}  

\begin{figure}   
\vskip 2.5 cm
    \includegraphics{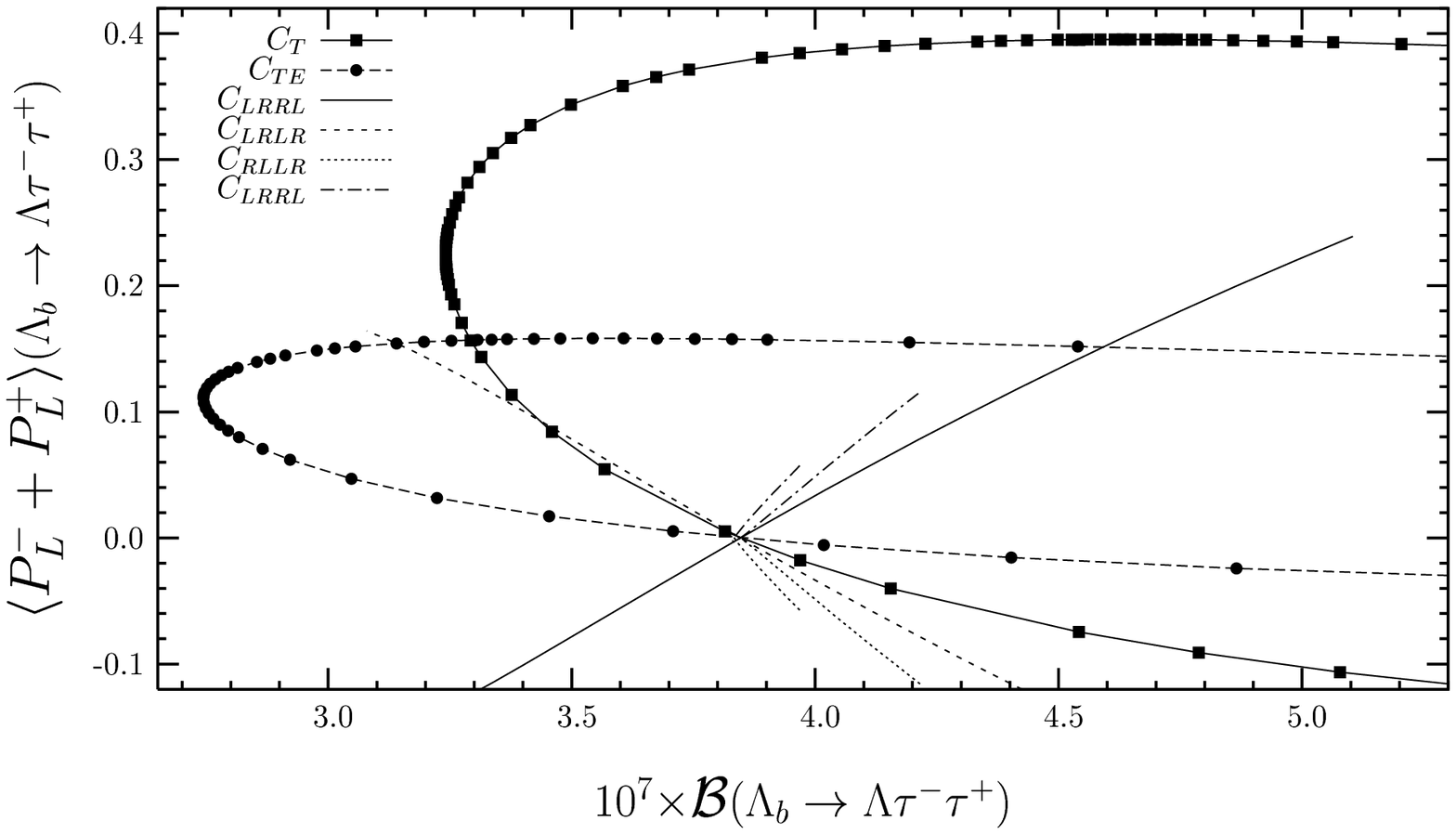}
\vskip 7.8 cm   
\begin{center}
{\bf Fig. 5}
\end{center}
\end{figure}

\begin{figure}   
\vskip 1.5 cm
    \includegraphics{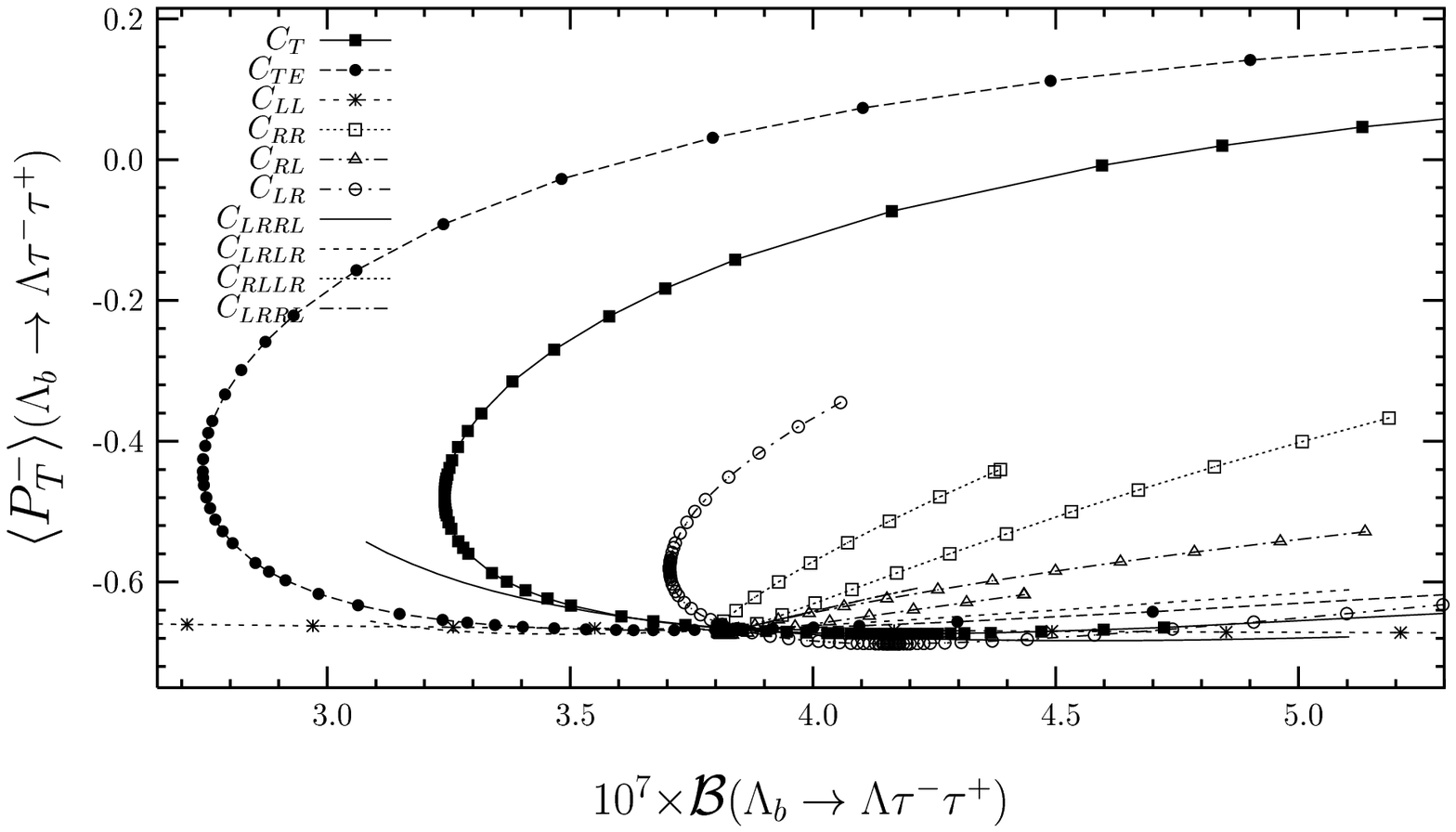}
\vskip 7.8cm
\begin{center}
{\bf Fig. 6}
\end{center}
\end{figure}

\begin{figure}    
\vskip 2.5 cm
    \includegraphics{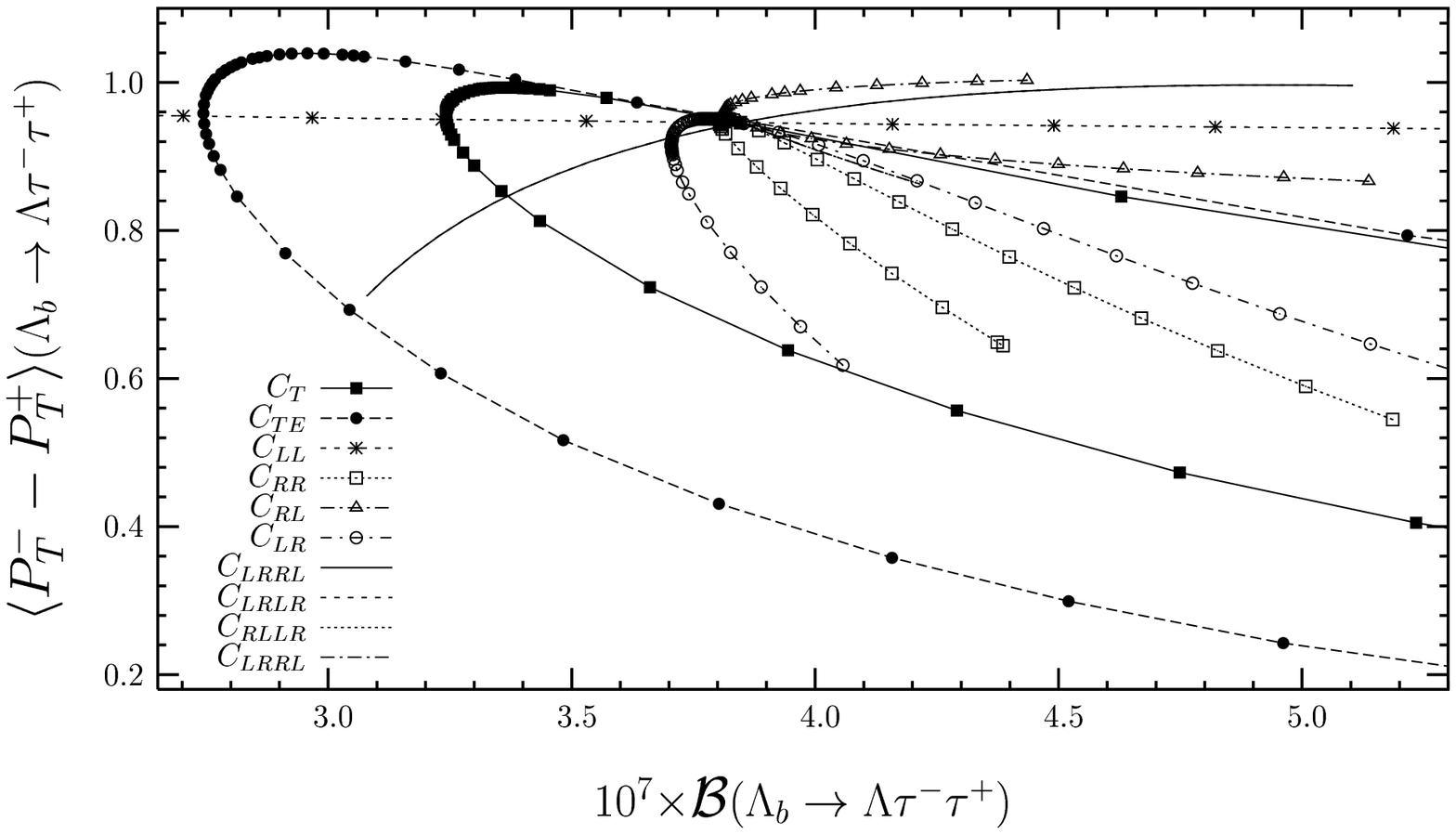}
\vskip 7.8 cm   
\begin{center}
{\bf Fig. 7}
\end{center}
\end{figure}

\begin{figure}
\vskip 1.5 cm
    \includegraphics{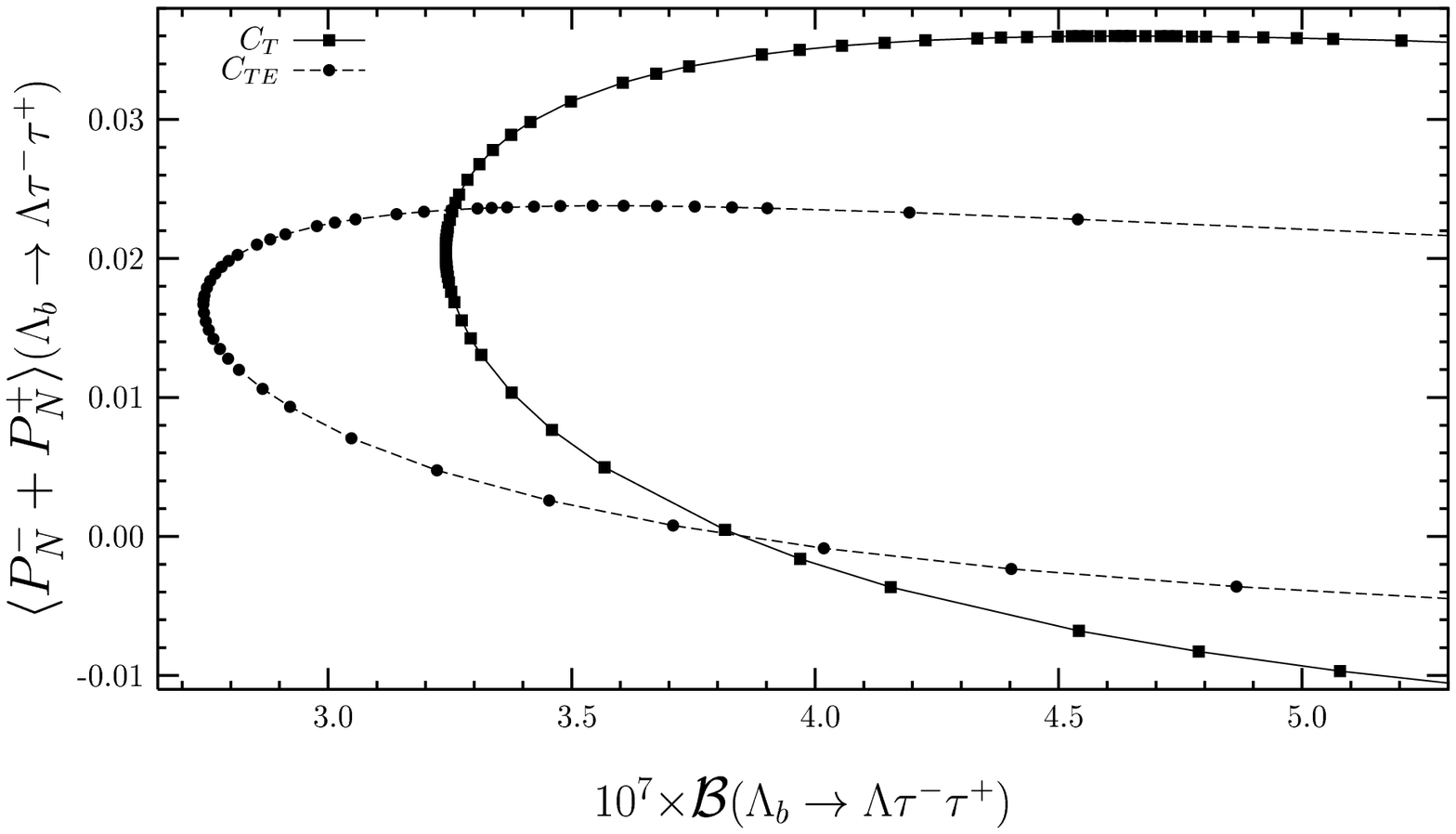}
\vskip 7.8cm
\begin{center}
{\bf Fig. 8}
\end{center}
\end{figure}

\end{document}